\documentclass{pacmem}
\usepackage{amssymb}

\begin{document}

\title{Cosmology of Rolling Tachyon
\thanks{Proceedings of Pac Memorial
Symposium on Theoretical Physics, pp.209--239 (Chungbum Publishing 
House, Seoul). Talk was given by Y. Kim.}}

\author{Chanju Kim}

\address{Department of Physics, Ewha Womans University, Seoul 120-750, Korea\\
E-mail: cjkim@ewha.ac.kr}

\author{Hang Bae Kim}

\address{Institute of Theoretical Physics, University of Lausanne,\\
CH-1015 Lausanne, Switzerland\\
E-mail: HangBae.Kim@ipt.unil.ch}

\author{Yoonbai Kim and O-Kab Kwon}

\address{BK21 Physics Research Division and Institute of Basic Science,\\
Sungkyunkwan University, Suwon 440-746, Korea\\
and\\
School of Physics, Korea Institute for Advanced Study,\\
207-43, Cheongryangri-Dong, Dongdaemun-Gu, Seoul 130-012, Korea\\ 
E-mail: yoonbai@skku.ac.kr,~okwon@newton.skku.ac.kr}


\maketitle

\abstracts{We study dynamics of rolling tachyon and Abelian
gauge field on unstable D-branes, of which effective action is given by
Born-Infeld type nonlocal action. Possible cosmological evolutions
are also discussed. In the Einstein frame of string cosmology, 
every expanding flat universe is proven to be decelerating.}

\section{Introduction}
Recent analysis of string theory~\cite{Gar,ST,GHY} suggests that classical dynamics of tachyon
on an unstable D-brane is described by a scalar Born-Infeld type action
with a runaway potential~\cite{GS}.
Characteristic of such tachyon effective theory is absence of perturbative
tachyonic modes around the potential minimum and existence of spatially
homogeneous rolling tachyon configuration~\cite{Sen}. In the background 
of the rolling tachyon, Abelian gauge field on the D-brane cannot support
perturbative spectrum at late-time~\cite{IU,MW,KKKK}. In this review, we 
summarize firstly bosonic effective action both in the bulk and on the 
D-brane, together with shapes of the tachyon potential. In section 3, 
we introduce Carrolian limit, and then discuss nonexistence of perturbative
spectra of the tachyon and the Abelian gauge field in the potential
minimum, of which dynamics are governed by nonlocal D-brane action.
For the tachyon, there exists the classical rolling tachyon solution
connecting smoothly unstable maximum of the potential and stable minimum 
at infinity.

Though the situation seems not to be matured yet, string cosmology must
be an intriguing subject to be tackled at every step of progress in both
string theory and observational cosmology~\cite{SC}.
Recently, inspired by string theory D-branes and heterotic M theory,
brane cosmology attracted much attention.
Brane cosmology assumes that our universe starts out with branes embedded
in the higher dimensional spacetime, either stable or unstable.
In such context, cosmology of the rolling tachyon must be an attractive 
subject~\cite{Gib,Ale,KL,tacos,KKK}. Topics included inflation, dark matter, 
cosmological perturbation, and reheating despite stringent difficulty 
in the simplest versions of this theory. A nice aspect for the cosmology
is nothing but the absence of open string excitations so that
classical analysis of late-time based on homogeneity and isotropy may
lead to a solid prediction. In section 4, we briefly describe 
system of gravity and the tachyon. The obtained universe in the simplest
set predicts that expanding universe is accelerating at onset stage but
turns to decelerating phase at late-time~\cite{Gib}. Finally, the
scale factor approaches a constant.
Our main concern is string cosmology involving graviton, dilaton, and the
tachyon. In the string frame, we obtain flat space solutions
at both the early epochs and late-times in closed form.
When transformed to the Einstein frame, analysis shows that every expanding
universe of the graviton-tachyon-dilaton system should be decelerating
irrespective of the specific shape of the tachyon potential. Specific form of
the scale factor is proportional to $(\mbox{time})^{1/2}$.

We conclude with a summary of this paper. Sometimes, same notations are used 
in different sections for convenience.

\section{Setup}
Effective action of bosonic sector of D3-brane system of tension $T_{3}$ 
is described by the  
following action in string frame~\cite{Pol,Gar}
\begin{eqnarray}
S &=& \frac{1}{2\kappa^2} \int d^4x\sqrt{-g}\;e^{-2\Phi} \left(
R +4\partial_\mu\Phi\partial^\mu\Phi
-\frac{1}{12}H_{\mu\nu\rho}H^{\mu\nu\rho} -2\Lambda
\right)  
\nonumber\\
&&
-T_3\int d^4x\;e^{-\Phi}V(T)
\sqrt{-\det(g_{\mu\nu} +
\partial_\mu T\partial_\nu T
+B_{\mu\nu}+
F_{\mu\nu})} \;, 
\label{act}
\end{eqnarray}
where cosmological constant $\Lambda$ is assumed to be vanished in what
follows.\footnote{Other effective actions different from Eq.~(\ref{act})
can also be considered. For example boundary string field theory(BSFT)
action for a non-BPS D-brane was also proposed and studied~\cite{ST}.}
The (quadratic) action in the first line of Eq.~(\ref{act}) 
is bulk action
of graviton $g_{\mu\nu}$, dilaton $\Phi$, and antisymmetric
tensor field $B_{\mu\nu}$ of rank 2 of which field strength $H_{\mu\nu\rho}$ 
is defined by $H_{\mu\nu\rho}=\partial_{\mu}B_{\nu\rho}+\partial_{\nu}
B_{\rho\mu}+\partial_{\rho}B_{\mu\nu}$. The action induced on the D3-brane
is Born-Infeld type nonlocal action given in the second line of 
Eq.~(\ref{act}), which involves kinetic terms of tachyon $T$ 
and Abelian gauge field $A_{\mu}$
satisfying $F_{\mu\nu}=\partial_{\mu}A_{\nu}-\partial_{\nu}A_{\mu}$.
Note that the tachyon $T$ and the Abelian gauge field $A_{\mu}$ were
rescaled as $T\rightarrow\sqrt{2\pi\alpha'}T$ and 
$A_{\mu}\rightarrow2\pi\alpha' A_{\mu}$ to absorb $2\pi\alpha'$ in front
of the kinetic terms.
 
Since we will deal with two classical topics in this paper, i.e., one is
string cosmology including the rolling tachyon~\cite{KKK} and 
the other is evolution of electromagnetic waves in the background of 
the rolling tachyon~\cite{KKKK}, let us firstly read equations of motion
from the action (\ref{act})
\begin{eqnarray}
R_{\mu\nu}-\frac12g_{\mu\nu}R &=& -\Lambda g_{\mu\nu}
-2\left[\nabla_\mu\nabla_\nu\Phi
    +g_{\mu\nu}\left(\nabla_\alpha\Phi\nabla^\alpha\Phi
        -\nabla_\alpha\nabla^\alpha\Phi\right)\right]
\nonumber\\ && 
+\left(\frac14H_{\mu\alpha\beta}H_\nu^{\alpha\beta}
   -\frac{1}{24}g_{\mu\nu}H_{\alpha\beta\gamma}H^{\alpha\beta\gamma}\right)
\nonumber\\ && 
+\kappa^2T_3\ e^\Phi\ V(T)\left(
\frac{g_{\mu\alpha}C^{\alpha\beta}\partial_\beta T\partial_\nu T}{\sqrt{gX}}
-g_{\mu\nu}\sqrt{g^{-1}X}\right),
\label{metric2}
\end{eqnarray}
\begin{equation}
4\left(-\nabla_\mu\nabla^\mu\Phi+\nabla_\mu\Phi\nabla^\mu\Phi\right) =
R -2\Lambda
-\frac{1}{12}H_{\alpha\beta\gamma}H^{\alpha\beta\gamma}
-\kappa^2T_3\;e^\Phi\;V(T)\sqrt{g^{-1}X},
\label{dilaton2}
\end{equation}
\begin{equation}
\nabla_\mu H^{\mu\nu\lambda}=2\nabla_\mu\Phi H^{\mu\nu\lambda},
\label{kr2}
\end{equation}
\begin{eqnarray}
&&C^{\mu\nu}_{(s)}\nabla_\mu \nabla_\nu T
-\frac{C^{\mu\nu}_{(s)}C^{\alpha\beta}\nabla_\mu T\nabla_\nu X_{\alpha\beta}
-X\nabla^\mu T\nabla_\nu C^{\mu\nu}_{(s)}}{2X}
-C^{\mu\nu}_{(s)}\nabla_\mu T\nabla_\nu\Phi \nonumber\\
&&\hspace{43mm}-\frac{1}{V}\frac{dV}{dT}\left(X-C^{\mu\nu}_{(s)}\nabla_\mu 
T\nabla_\nu T\right) =0,
\label{tachyon2}
\end{eqnarray}
where
$X_{\mu\nu}=g_{\mu\nu}+B_{\mu\nu}+F_{\mu\nu}+\partial_\mu T\partial_\nu T$,
$X=\det(X_{\mu\nu})$, $C^{\mu\nu}$ is the cofactor of $X_{\mu\nu}$,
and $C^{\mu\nu}_{(s)}$ is the symmetric part of $C^{\mu\nu}$.

\begin{figure}
\centerline{\epsfig{figure=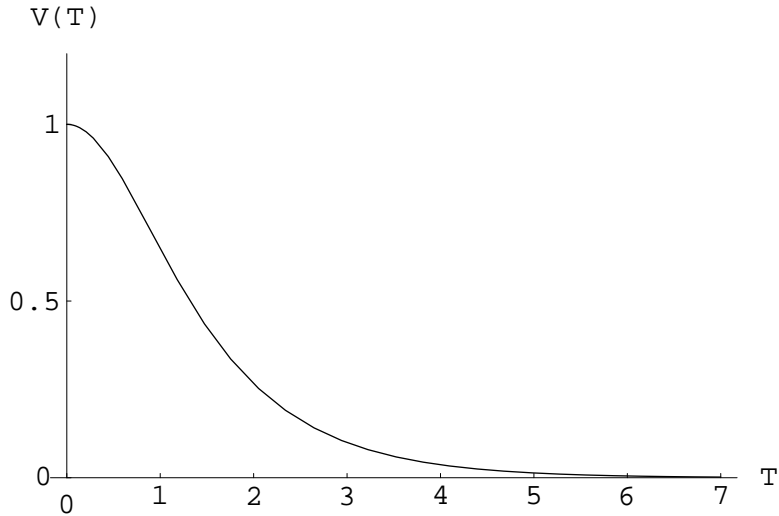,height=7cm}}
\caption{
Tachyon potential ${\displaystyle V(T)=\frac{1}{{\rm cosh}T}}$
}
\label{fig1}
\end{figure}

The tachyon potential $V(T)$ multiplied in front of the Born-Infeld type action 
(\ref{act}) is known to measure varying tension of the unstable D-branes.
Though there seem no consensus and no exact computation on the
form of tachyon potential $V(T)$, it should have its maximum, max($V(T))=1$, 
and minimum, min($V(T))=0$. Some known computations from string theories are
as follows~\cite{GS}. For bosonic string theory, we have
\begin{equation}\label{V1}
V(T)=\left\{
\begin{array}{ll}
\frac{T^{2}}{8}\exp\left(1-\frac{T^{2}}{8}\right) & \mbox{for small $T$
but $T\ge 1$} \\
\exp\left(-\frac{T}{2}\right) & \mbox{for large $T$} 
\end{array}
,
\right.
\end{equation}
and, for superstring theory, we have
\begin{equation}\label{V2}
V(T)=\left\{
\begin{array}{ll}
\exp\left(-\frac{T^{2}}{8\ln 2}\right) & \mbox{for small $T$ but $T\ge 0$} \\
\exp\left(-\frac{T}{\sqrt{2}}\right) & \mbox{for large $T$}
\end{array}
.
\right.
\end{equation}
String theory calculation (\ref{V1})--(\ref{V2}) tells us that 
favorable form of the tachyon potential supports $V\sim e^{-T^{2}}$ 
for small $T$ and $V\sim e^{-T}$ for large $T$.
Since the potential $V(T)$ must be a smooth function in the intermediate region,
connecting two asymptotic expressions, a typical example is~\cite{KL}
\begin{equation}\label{V3}
V(T)=\left(1+\frac{T}{T_{0}}\right)\exp\left(-\frac{T}{T_{0}}\right).
\end{equation}
In this paper, considering convenience in analytic computation, we adopt
\begin{equation}\label{V4}
V(T)=\frac{1}{{\rm cosh}\left(\frac{T}{T_{0}}\right)}
\end{equation}
which is also consistent with the results of string theory 
(\ref{V1})--(\ref{V2}) (See Fig.~\ref{fig1}).

\section{Tachyon and Gauge Field}

Among various bosonic degrees of freedom, dynamics of the tachyon $T$ and
the gauge field $A_{\mu}$ is governed by nonlocal kinetic terms 
in the brane action (\ref{act}).
Before discussing their dynamics directly, let us consider a massless 
noninteracting real scalar field $\phi$ described by Lorentz-invariant 
linear wave equation
\begin{equation}
\left(\frac{1}{v^{2}}\frac{\partial^{2}}{\partial t^{2}}-\nabla^{2}\right)
\phi=0.
\end{equation} 
If we take two singular limits known as the In\"{o}n\"{o}-Wigner limits,
infinite velocity limit, $v\rightarrow \infty$, goes to Galileo's limit
of action at a distance described by an elliptic equation~:
\begin{equation}
\nabla^{2}\phi=0
\end{equation} 
which we encounter frequently in Newtonian dynamics. On the other hand, 
the opposite velocity dominated limit, $v\rightarrow 0$, called as
Carrolian limit is described by an ordinary differential equation~:
\begin{equation}\label{carr}
\frac{d^{2}}{dt^{2}}\phi=0.
\end{equation}

A natural implication from the above discussion is possible absence of plane
waves in the Carrolian limit of corresponding field theory. Instead 
time evolution of homogeneous configuration may be dominated. 
Though both limits are singular in the context of group theory, we will
use the speed $v$ as a continuous parameter of range $(0\le v\le 1)$ as we 
have done in the Galileo's limit where its value is taken from 
from one to infinity.

In this section, we will briefly review rolling tachyon in flat spacetime
with disappearance of perturbative tachyonic spectrum at the bottom of
the potential $V(T)$ and discuss singular evolution of perturbative 
electromagnetic field in background of the rolling tachyon.

\subsection{Rolling tachyon in flat spacetime}

Let us begin this subsection with discussing physics of unstable D-branes 
in the simplest setup. If we consider solely motion of the tachyon 
in flat spacetime, it is depicted by~\cite{Sen} 
\begin{equation}\label{tact}
S_{T}=-T_{3}\int d^{4}x V(T)\sqrt{1+\partial_{\mu}T\partial^{\mu}T},
\end{equation}
where specific form of the tachyon potential $V(T)$ is chosen by 
Eq.~(\ref{V4}). 
Note that $g^{-1}\det(g_{\mu\nu}+\partial_\mu T\partial_\nu T)=
1+\partial_\mu T\partial^\mu T$ up to the quadratic terms
and it holds in an exact form as far as homogeneous configurations are
concerned such as the cosmology described in terms of
Robertson-Walker metric of our interest. Equation of motion is
\begin{equation}\label{taeq1}
\frac{1}{\sqrt{-\bar{\eta}}}\partial_{\mu}\left(V\sqrt{-\bar{\eta}}
\bar{\eta}^{\mu\nu}\partial_{\nu}T\right)=\frac{dV}{dT},
\end{equation}
where $\bar{\eta} = -(1 + \partial_\mu T \partial^\mu T)$ and 
\begin{eqnarray}\label{beta}
\bar{\eta}_{\mu\nu}=\eta_{\mu\nu}+\partial_{\mu}T\partial_{\nu}T,
\qquad
\bar{\eta}^{\mu\nu} = \eta^{\mu\nu} -
\frac{\partial^\mu T \partial^\nu T}{1
+ \partial_\rho T \partial^\rho T}.
\end{eqnarray}
Energy-momentum tensor $T_{\mu\nu}$ is given by
\begin{eqnarray}\label{temt}
T_{\mu\nu} = -T_3  V(T) \eta_{\mu\rho}\eta_{\nu\sigma}\sqrt{-\bar{\eta}}
\bar{\eta}^{\rho\sigma}.
\end{eqnarray}

Since $T=0$ is unique stationary point of the tachyon potential $V(T)$, 
we perturb
the action (\ref{tact}) up to quadratic terms to read perturbative spectrum.
For small $T$, we have
\begin{equation}
S_{T}+T_{3}\int d^{4}x \approx 
T_{3}\int d^{4}x\left(-\frac{1}{2}\partial_{\mu}T
\partial^{\mu}T+\frac{1}{2T_{0}^{2}}T^{2}\right),
\end{equation}
where the positive cosmological constant at $T=0$ is subtracted in the action.
As expected, the sign in front of mass term is negative so it is tachyonic
on top of the potential $V(T=0)$. For large $T$, the tachyon potential 
is approximated as $V(T)\approx 2e^{-T/T_{0}}$. Introducing a new small
order parameter such as $\phi=2\sqrt{2}T_{0}e^{-T/2T_{0}}$, we obtain a 
quadratic action with positive mass term
\begin{equation}\label{disac}
S_{T}\approx T_{3}\int d^{4}x \left[-\frac{1}{2}\partial_{\mu}\phi
\partial^{\mu}\phi-\frac{1}{2}\left(\frac{1}{\sqrt{2}T_{0}}\right)^{2}
\phi^{2} \right].
\end{equation}  
Though this seems to imply existence of a perturbative tachyonic 
spectrum of mass 
$1/\sqrt{2}T_{0}$, it is an artifact of quadratic approximation since
$T\rightarrow \infty$ is not a stationary point of the tachyon potential
as shown in the Fig.~\ref{fig1}. 
To be specific, let us consider a plane wave solution
$\phi\sim e^{ik_{\mu}x^{\mu}}$ and then we have 
$\partial_{\mu}T\sim -2iT_{0}k_{\mu}$. Inserting these into the equation 
of motion (\ref{taeq1}), we get the following dispersion relation
\begin{equation}\label{disp}
\frac{1}{\sqrt{1-4T^{2}_{0}k^{2}}}\frac{d\ln V}{dT}=0.
\end{equation}
Note that $d(\ln V)/dT|_{T\rightarrow \infty}=-T_{0}\ne 0$, the dispersion
relation does not have any nontrivial solution but can only be satisfied 
for infinite $k^{2}$. The mass-shell condition we derived through a
quadratic approximation (\ref{disac}) was reproduced by a Taylor expansion of 
Eq.~(\ref{disp}) up to quadratic terms for small $k^{2}$.

Absence of perturbative tachyon spectrum does not lead to nonexistence of 
classical configurations. We may take into account an intriguing lesson
from the Carrolian limit (\ref{carr}) of the wave equation. Suppose that 
we have a homogeneous tachyon configuration with $\partial_{i}T=0$ and 
consider its classical time evolution $T(t)$. Then only diagonal components of
the energy-momentum tensor (\ref{temt}) survive and the equation of motion 
(\ref{taeq1}) is integrated to conservation of energy 
$\rho_{T}\equiv T_{00}$, i.e., 
Hamiltonian density ${\cal H}_T$ is
\begin{equation}\label{taeq}
{\cal H}_T=\rho_{T}
=T_{3}\frac{V(T)}{\sqrt{1-\dot{T}^{2}}}=T_{3}\epsilon,
\end{equation}
where $\epsilon=V(T_i)/\sqrt{1-\dot T_i^2}\ge 1$ and
$T_i$, $\dot T_i$ are the initial values of $T$ and $\dot T$.
It is analogous to the aforementioned Carollian limit (\ref{carr}).
Constancy of the Hamiltonian density (\ref{taeq}) forces $\dot{T}\rightarrow
1$ for any finite $\epsilon$ as $T\rightarrow \infty$.
For the tachyon potential (\ref{V4}), we have a solution of rolling tachyon
\begin{eqnarray}\label{ts1}
T(t) = T_0\sinh^{-1}\left(ae^{t/T_0}-\frac{b}{4a}e^{-t/T_0}\right),
\end{eqnarray}
where $a=\frac12\left[\sinh(T_i/T_0)+\sqrt{\sinh^2(T_i/T_0)+b}\right]$
and $b=(\epsilon^2-1)/\epsilon^2$.
At late-time regime, the obtained tachyon solution (\ref{ts1})
is approximated as expected, i.e., $T(t)\approx t$ irrespective of the 
initial values of $T$ and $\dot{T}$.
Note that the pressure $p_{T}\equiv T^{i}_{\;i}/3$ is negative and 
approaches zero as time elapses, which
coincides with vanishing Lagrangian limit
\begin{equation}\label{press}
p_{T}={\cal L}_{T}=-T_{3}V(T)\sqrt{1-\dot{T}^{2}}
\stackrel{t\rightarrow\infty}{\longrightarrow}0.
\end{equation}
{}From equation of state $p_{T}=w_{T}\rho_{T}$, we read
\begin{eqnarray}\label{rw1}
w_{T}=-(1-\dot{T}^{2})=-\frac{V(T(t))^{2}}{\epsilon^{2}}\le 0,
\end{eqnarray}
where $w_{T}=-1/\epsilon^{2}$ at $T=0$ and $w_{T}\rightarrow 0$ at 
$T\rightarrow \infty$.

\subsection{Gauge field and rolling tachyon}
When we turn off antisymmetric tensor field of second rank $B_{\mu\nu}$
in the Born-Infeld type action in Eq.~(\ref{act}),
effective bosonic action of unstable D3-brane system is given 
by~\cite{IU,MW,GHY,RS,KKKK} 
\begin{equation}\label{fa}
S= -T_3 \int d^4x\; V(T) \sqrt{-\det (\eta_{\mu\nu} +
\partial_\mu T\partial_\nu T + F_{\mu\nu})}\, .
\end{equation}
To proceed, we introduce a few notations. We first define
\begin{eqnarray}
X_{\mu\nu}&\equiv & \eta_{\mu\nu} + \partial_\mu T\partial_\nu T + F_{\mu\nu},
\label{Xmn}\\
X&\equiv & \det (X_{\mu\nu}),
\label{X}
\end{eqnarray}
and barred covariant field strength tensor $\bar{F}_{\mu\nu}$
\begin{eqnarray}\label{fmn}
\bar{F}_{\mu\nu}=F_{\mu\nu} 
\end{eqnarray}
with Eq.~(\ref{beta}).
Then we have contravariant barred field strength tensor $\bar{F}^{\mu\nu}$ and
its dual field strength $\bar{F}^{\ast}_{\mu\nu}$
\begin{eqnarray}\label{ffmn}
\bar{F}^{\mu\nu} = \bar{\eta}^{\mu\alpha}
\bar{\eta}^{\nu\beta}F_{\alpha\beta},~~~
\bar{F}^{\ast}_{\mu\nu}= \frac{\bar{\epsilon}_{\mu\nu\alpha\beta}}
{2}
\bar{F}^{\alpha\beta}= \frac{\bar{\epsilon}_{\mu\nu\alpha\beta}}{2}
\bar{\eta}^{\alpha\gamma}\bar{\eta}^{\beta\delta}F_{\gamma\delta},
\end{eqnarray}
where $ \bar{\epsilon}_{\mu\nu\alpha\beta} = \sqrt{-\bar{\eta}}\;
\epsilon_{\mu\nu\alpha\beta}$ with $\epsilon_{0123} = 1$.

In terms of barred quantities
Eq.~(\ref{X}) is computed as
\begin{eqnarray}\label{XX}
X = \bar{\eta} \left[ 1 + \frac12 \bar{F}_{\mu\nu}
\bar{F}^{\mu\nu} - \frac1{16} \left(\bar{F}^*_{\mu\nu}
\bar{F}^{\mu\nu}\right)^2 \right].
\end{eqnarray}
Then equations of motion for the tachyon $T$ and the gauge field
$A_{\mu}$ are
\begin{eqnarray}
\partial_\mu\left( \frac{V}{\sqrt{-X}}
\;C^{\mu\nu}_{\rm S}\; \partial_\nu T\right)
+\sqrt{-X}\; \frac{d V}{d T}
= 0,
\label{te} \\
\partial_\mu\left( \frac{V}{\sqrt{-X}}
\;C^{\mu\nu}_{\rm A}\right) = 0.
\label{ge}
\end{eqnarray}
Here $C^{\mu\nu}_{\rm S}$ and $C^{\mu\nu}_{\rm A}$ are symmetric
and asymmetric part, respectively, of the cofactor,
\begin{equation}\label{cmn}
C^{\mu\nu} = \bar{\eta}\left(
\bar{\eta}^{\mu\nu} + \bar{F}^{\mu\nu} + \bar{\eta}^{\mu\alpha}
\bar{\eta}^{\beta\gamma}\bar{\eta}^{\delta\nu}
\bar{F}^*_{\alpha\beta}\bar{F}^*_{\gamma\delta}
+\bar{\eta}^{\mu\alpha}\bar{\eta}^{\beta\gamma}
\bar{F}^*_{\alpha\beta}\bar{F}^*_{\gamma\delta}\bar{F}^{\delta\nu}
\right),
\end{equation}
namely,
\begin{eqnarray}
C^{\mu\nu}_{\rm S} &=& \bar{\eta} (
\bar{\eta}^{\mu\nu}
+ \bar{\eta}^{\mu\alpha}
\bar{\eta}^{\beta\gamma}\bar{\eta}^{\delta\nu}
\bar{F}^*_{\alpha\beta}\bar{F}^*_{\gamma\delta}),\\
C^{\mu\nu}_{\rm A} &=& \bar{\eta}(\bar{F}^{\mu\nu}
+\bar{\eta}^{\mu\alpha}\bar{\eta}^{\beta\gamma}
\bar{F}^*_{\alpha\beta}\bar{F}^*_{\gamma\delta}\bar{F}^{\delta\nu}).
\end{eqnarray}

Energy-momentum tensor $T_{\mu\nu}$ is given by
\begin{eqnarray}
T_{\mu\nu} = - \frac{T_3 V(T)}{\sqrt{- X}}\;
\left[ -\eta_{\mu\nu} X
+  C_{\mu\rho}(\partial_\nu T \partial^\rho T
+ F_\nu^{~\rho})\right],
\end{eqnarray}
where $C_{\mu\nu}\equiv \eta_{\mu\alpha}\eta_{\nu\beta}
C^{\alpha\beta}$.
The diagonal components can be identified as density and pressure,
\begin{eqnarray}
\rho&=&  - \frac{T_3 V(T)}{\sqrt{- X}}\;
\left[ X +  C_{0\mu} (\dot{T}\partial^\mu T
+ F_0^{~\mu} )\right],  \label{rd}\\
p &=&  - \frac{T_3 V(T)}{3\sqrt{- X}}\;
\left[-3 X + C_{0\mu}(\dot{T}\partial^\mu T + F_0^{~\mu})
 + C^{\mu\nu}(\partial_\mu T\partial_\nu T + F_{\mu\nu})
\right]. \label{rp}
\end{eqnarray}
{}From equation of state $p=w \rho$,
we read
\begin{eqnarray}
w &=& \frac13 - \frac{4 X - C^{\mu\nu}
(\partial_\mu T\partial_\nu T + F_{\mu\nu})}
{ 3\left[ X + C_{0\mu}(\dot{T}\partial^\mu T + F_0^{~\mu})
\right]}.\label{rw}
\end{eqnarray}
Thus the trace of the total energy-momentum tensor is given by
\begin{equation}\label{trt}
T^\mu_{~\mu} = -\frac {T_3 V(T)}{\sqrt{-X }} \left[ (4 + 3
\partial_\rho T \partial^\rho T) +\left((1+\frac12 \partial_\rho T
\partial^\rho T)\delta^\mu_{~\sigma} - \partial^\mu T \partial_\sigma T
\right) F_{\mu\nu} F^{\sigma\nu} \right].
\end{equation}

The equations of motion (\ref{te}) and (\ref{ge}) admit homogeneous solutions
$T(t)$ and $A_\mu(t)$ which satisfy
\begin{eqnarray}
&& \partial_0 \left( \frac{V}{\sqrt{-X}} \dot{T} \right)
  + \sqrt{-X} \frac{dV}{dT} = 0, \nonumber \\
&& \partial_0 \left( \frac{V}{\sqrt{-X}} E^i \right) = 0,
\end{eqnarray}
where $X = -1 + \dot{T}^2 + {\bf E}^2$. Solving these equations we find that
$V/\sqrt{-X} = \mbox{constant}$ and ${\bf E} = \mbox{constant}$.
The first equation is nothing but the statement of conservation of energy.

When ${\bf E} = 0$, the solution reduces to that in the theory
with pure tachyon field given in the subsection 3.1.
When ${\bf E} \neq 0$, one can simply read the solution from Eq.~(\ref{ts1})
by rescaling $\epsilon^2\ \rightarrow \epsilon^2(1-E^2)$
and $T_0 \rightarrow T_0/\sqrt{1-E^2}$ in the exponents of Eq.~(\ref{ts1}).

Now we consider the propagation of the gauge field $A_{\mu}$ in the
rolling tachyon background (\ref{ts1}). For this purpose we linearize the
equation of motion (\ref{ge}) in this background. Written in terms
of electric and magnetic fields, Amp\'{e}re's law is modified as
\begin{equation}\label{ameq}
\nabla\times {\bf B}=\frac{1}{V\sqrt{1-\dot{T}^{2}}}
\frac{\partial}{\partial t}\left(\frac{V}{\sqrt{1-\dot{T}^{2}}}{\bf E}\right).
\end{equation}
Combined with the Bianchi identity, the wave wave equations of the
electric field ${\bf E}$ and the magnetic field ${\bf B}$ become
\begin{eqnarray}
\frac{1}{V\sqrt{1-\dot{T}^{2}}}\frac{\partial}{\partial t}
\left(\frac{V}{\sqrt{1-\dot{T}^{2}}}\frac{\partial {\bf B}}{\partial t}\right)
-\nabla^{2}{\bf B}&=&0,\label{beq}\\
\frac{\partial}{\partial t}\left[\frac{1}{V\sqrt{1-\dot{T}^{2}}}
\frac{\partial}{\partial t}\left(\frac{V{\bf E}}{\sqrt{1-\dot{T}^{2}}}
\right)\right] -\nabla^{2}{\bf E}&=&0.\label{eeq}
\end{eqnarray}
Note that these equations are not identical.
Since the Hamiltonian density ${\cal H}_{T}$ (\ref{taeq}) 
is a constant of motion, 
the wave equations (\ref{beq}) and (\ref{eeq}) reduce to
\begin{eqnarray}
\frac{\epsilon^2}{V^2}\frac{\partial^2{\bf B}}{\partial t^2}
-\nabla^2{\bf B} &=& 0,\label{beq1}\\
\frac{\epsilon^2}{V^2}\frac{\partial^2{\bf E}}{\partial t^2} +
\frac{d}{dt}\left(\frac{\epsilon^2}{V^2}\right)
\frac{\partial{\bf E}}{\partial t}
-\nabla^2{\bf E} &=& 0. \label{eeq2r}
\end{eqnarray}
One can also compute
the energy density (\ref{rd}) and pressure (\ref{rp}) as
are given by
\begin{eqnarray}
\rho_{{\rm em}} &\equiv& T_{00} - T_{00}|_{F_{\mu\nu}=0} 
\approx \frac{T_3}2 \frac{V(T)}{\sqrt{1 - \dot{T}^2}} \left(
\frac{{\bf E}^2}{1-\dot{T}^2} +
 {\bf B}^2\right), \label{eden}\\
p_{{\rm em}} &\equiv&\frac13(T_{ii} - T_{ii}|_{F_{\mu\nu}=0}) 
\approx \frac{T_3}{6}\frac{ V(T)}{ \sqrt{1-\dot{T}^2}} \left[ 
 {\bf E}^2 +(1- \dot{T}^2 ) {\bf B}^2 \right].\label{pre1}
\end{eqnarray}
Similarly, we read $w$ from Eq.~(\ref{rw})
\begin{equation}
w_{{\rm em}} \equiv \frac{p_{{\rm em}}}{\rho_{{\rm em}}}\approx \frac{1}{3} (1-\dot{T}^2). 
\label{eos1}
\end{equation}
The trace of the energy-momentum tensor (\ref{trt}) becomes
\begin{equation}
T^{~~\mu}_{em~\mu}\equiv T^\mu_{~\mu} - T^\mu_{~\mu}|_{F_{\mu\nu}=0}\approx 
-\frac{T_3}2\frac{V(T)}{\sqrt{1-\dot{T}^2}}\dot{T}^2
\left( \frac{{\bf E}^2}{1-\dot{T}^2} +  {\bf B}^2 \right).
\label{tmm1}
\end{equation}

\begin{figure}
\centerline{\epsfig{figure=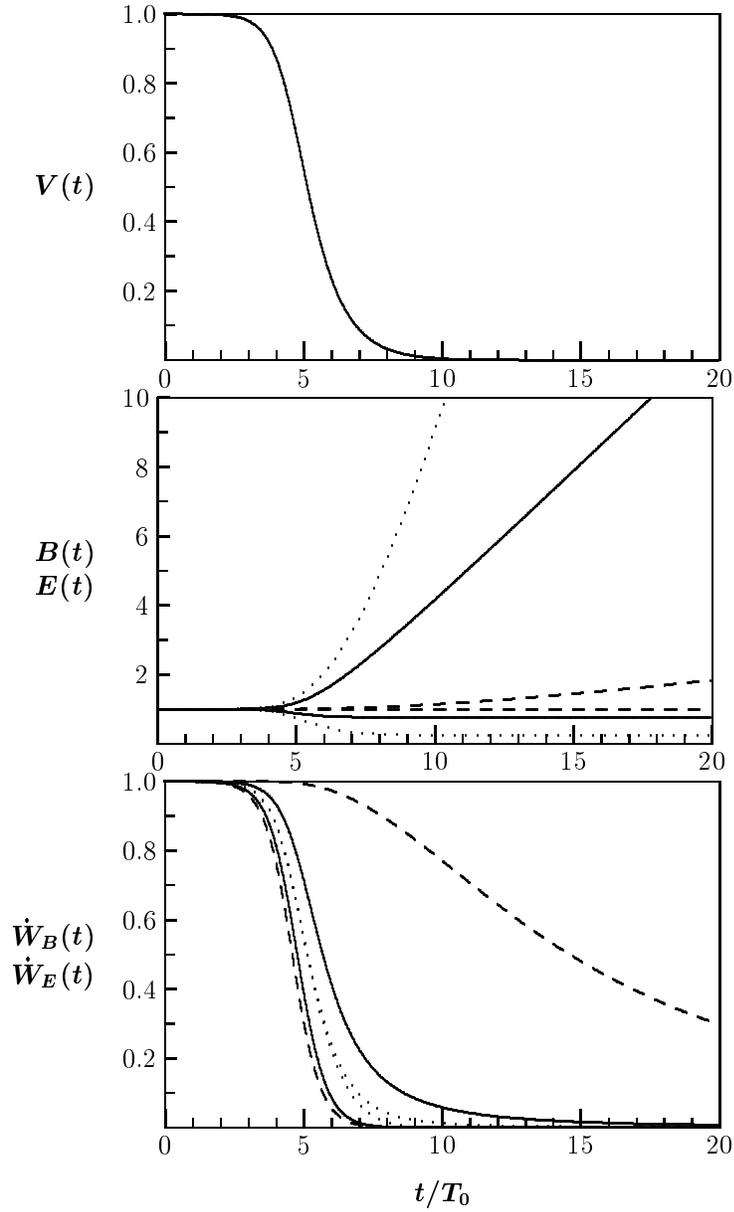,height=16cm}}
\caption{
(Top) tachyon potential as a function of time, Eq.~(\ref{vtt})
for $a=0.1$ and $b=0$ ($\epsilon=1$).
(Middle) amplitudes and (bottom) propagation speeds of
magnetic field (upper three lines) and electric field (lower three lines)
for $kT_0=0.1$ (dashed line), $kT_0=1$ (solid line),
and $kT_0=10$ (dotted line).
}
\label{fig2}
\end{figure}

The speed of wave propagation is altered by $V(T(t))$ and thus time-varying.
One might expect from Eq.~(\ref{beq1}) that the propagation speed
roughly follows $V(t)$, but this is not necessarily so as we will see below.
Considering the shape of the potential in Eqs.~(\ref{V1}) and (\ref{V2}),
we see that it is the light speed on the top of the potential and
decreases monotonically to zero as the rolling of the tachyon proceeds.
For illustrative purpose, we take the tachyon potential (\ref{V3})
and the solution (\ref{ts1}).
Inserting the solution into the tachyon potential,
we have an expression of $V$ as a function of time $t$
\begin{equation}\label{vtt}
V(T(t)) =\frac{1}{\sqrt{1+\left(ae^{t/T_0}-\frac{b}{4a}e^{-t/T_0}\right)^2}}.
\end{equation}
As shown in Fig.~\ref{fig2},
$V(T(t))$ changes in the scale of a few $T_0$'s from the light speed to zero,
which is nothing but elapsed time from the onset of the tachyon rolling
to the late-time.
Since the electric field and the magnetic field obey different wave equations,
they show different behaviors in propagation.

To investigate in detail the propagation of electromagnetic wave in the
rolling tachyon background,
let us try to find a plane-wave solution of the wave equations with
a time-varying propagation speed, (\ref{beq1}) and (\ref{eeq2r}).
We consider the plane-wave propagating along the $x$-axis,
${\bf k}=k\hat{\bf x}$, and make an ansatz ${\bf B}=B_z(t,x)\hat{\bf z}$ with
\begin{equation} \label{Ba}
B_z(t,x) = B(t)e^{ik[x-W_B(t)]},
\end{equation}
where $B(t)$ and $W_B(t)$ are real.
We assume that $V(0)\approx1$ and $\dot W_B(0)\approx1$ so that
the ansatz (\ref{Ba}) gives the normal plane-wave initially.
Since $\bf E$ and $\bf B$ are related through the Amp\'ere's law
(\ref{ameq}) and Faraday's law
\begin{equation}\label{faraday}
\nabla\times{\bf E} + \frac{\partial{\bf B}}{\partial t} = 0,
\end{equation}
the electric field ${\bf E} = E_y(t,x) \hat{\bf y}$ is then given by
\begin{equation}
E_y(t,x) = E(t)e^{ik[x-W_E(t)]},
\end{equation}
with
\begin{eqnarray}
E &=& \sqrt{\dot W_B^2B^2+k^{-2}\dot B^2}, \label{e-b1}\\
W_E &=&  W_B - k^{-1}\tan^{-1}\left(\frac{k^{-1}\dot B}{\dot W_BB}\right).
\label{e-b2}
\end{eqnarray}
With this ansatz, Eq.~(\ref{beq1}) reduces to
\begin{eqnarray}
\ddot B-k^2\left(\dot W_B^2-\epsilon^{-2}V^2\right)B &=& 0, 
\label{fir}\\
2\dot B\dot W_B + B\ddot W_B &=& 0. \label{sec}
\end{eqnarray}
Eq.~(\ref{sec}) implies
\begin{equation}
\dot W_B = \frac{1}{\epsilon}\left(\frac{B_0}{B}\right)^2, \label{Bspeed}
\end{equation}
where $B_0$ is a constant. From Eq.~(\ref{e-b2}) we also find
\begin{equation} \label{Espeed}
\dot W_E = \frac{V^2}{\epsilon^3}\left(\frac{B_0}{E}\right)^2.
\end{equation}
Inserting Eq.~(\ref{Bspeed}) into Eq.~(\ref{fir}), we obtain a
second-order nonlinear equation for amplitude of the magnetic field
\begin{equation}
\ddot B - \frac{k^2}{\epsilon^2}\left[
   \left(\frac{B_0}{B}\right)^4 - V^2 \right] B = 0. 
\label{Beq}
\end{equation}
With the knowledge of $V(T(t))$, like the one given in Eq.~(\ref{vtt}),
we can solve these equations for $B(t)$ and $E(t)$.
For $V(T(t))$ of Eq.~(\ref{vtt}), the numerical solution is shown
in Fig.~\ref{fig2}.
Let us first examine a few limiting cases.

On the top of the potential $V=1$, we get a normal plane-wave solution
$\dot W_B=\dot W_E=1/\epsilon$ with amplitudes of $B = B_0$ and
$E=B_0/\epsilon$ as expected from the original wave equations
(\ref{beq1})--(\ref{eeq2r}).
At the top of the potential, the equation of states (\ref{eos1})
gives $w_{{\rm em}} = 1/3$ and the trace of the energy-momentum tensor
$T^{~~\mu}_{{\rm em}~\mu}$ up to quadratic terms (\ref{tmm1}) vanishes.

Once the tachyon starts
rolling from the top of the potential, the fluctuations of the gauge fields
starts deviating from the normal plane wave solution. With the
initial condition $V(t=0)=1$ for which $4a^2=b$, $V(T(t))$ in Eq.~(\ref{vtt})
reduces to
\begin{equation}
V(T(t)) = \frac{1}{\sqrt{1 + b \sinh^2 t/T_0}}.
\end{equation}
Assuming that $b = (\epsilon^2-1)/\epsilon^2$ is small, which means
the tachyon rolls slowly initially, we can solve 
Eqs.~(\ref{Beq}) and (\ref{Bspeed})
to the first order in $b$:
\begin{eqnarray}
B(t) &=& B_0 \left[1 + \frac{bk^2 T_0^2}{8(1+k^2 T_0^2)} \sinh^{2}t/T_0 
   \right], \nonumber \\
W_B(t) &=& \frac{t}{\epsilon} + \frac{b}{4\epsilon}\left[
     t - \frac{k^2 T_0^3}{2(1+k^2T_0^2)}\sinh{2t/T_0} \right], 
\end{eqnarray}
where we rescaled $B_0$ so that $B(0) = B_0$. This solution is valid
as long as $b e^{t/T_0} \ll 1$.
Thus the amplitude initially grows quadratically as the tachyon starts rolling.
On the other hand, electric field can be calculated from (\ref{e-b1}) and
(\ref{Espeed}),
\begin{eqnarray}
E(t) &=& B_0 \left[1 - \frac{b}{4(1+k^2 T_0^2)}
   \left(1+2k^2T_0^2 + k^2 T_0^2 \sinh^2{t/T_0}\right) \right], \nonumber \\
W_E(t) &=& \frac{t}{\epsilon} - \frac{b}{8\epsilon(1+k^2T_0^2)}\left[
     2 k^2T_0^2 t + (2+k^2T_0^2)\sinh{2t/T_0} \right]. 
\end{eqnarray}
Note that $E(t)$ and $\dot W_E$ tends to decrease in contrast with the
behavior of magnetic field. This can be confirmed in Fig.~\ref{fig2}.
As time elapses, $w_{{\rm em}}$ in Eq.~(\ref{eos1}) starts decreasing 
but $T^{~~\mu}_{{\rm em}~\mu}$
up to quadratic terms in Eq.~(\ref{tmm1}) develops nonvanishing
contributions.

At late-time, the value of the 
potential (\ref{vtt}) decays exponentially to zero
\begin{equation}
V(T(t)) \approx \frac{1}{\sqrt{b} e^{t/T_0}}.
\end{equation}
In this case, it is more convenient to work with the original equation
(\ref{beq1}) from which we see that the time dependence is exponentially
suppressed at late time. Therefore the magnetic field linearly diverges
in general,
\begin{equation}\label{Beql}
B(t) = B_1 t + B_2 + O(e^{-t/T_0}).
\end{equation}
Then the propagation speed decreases
according to Eq.~(\ref{Bspeed}) (See also Fig.~\ref{fig2}).
Since $B$ linearly increases, our quadratic approximation of the Born-Infeld
action is not valid as $t$ gets large. However, one may still
extract some relevant physics as the tachyon rolls to the vacuum,
$V\rightarrow0$. The electromagnetic waves become frozen and stop propagating.
For the magnetic field, short wavelengths freeze earlier than long wavelengths.
For the electric field in this limit, we have from Eq.~(\ref{faraday}),
\begin{equation}
E(t) = B_1/k,
\end{equation}
which is a constant. Note that in (\ref{Espeed}) $\dot W_E$ contains
extra $V^2$ compared with $\dot W_B$ and hence the electric field is
frozen more quickly than the magnetic field.

The density $\rho_{{\rm em}}$ (\ref{eden}) diverges 
exponentially from the electric
component but the pressure $p_{{\rm em}}$ (\ref{pre1}) 
approaches a finite value, 
so that $w_{{\rm em}}$ in Eq.~(\ref{eos1}) becomes zero. 
Similarly, the trace of 
the energy-momentum tensor (\ref{tmm1}) also diverges exponentially.
Since the Hamiltonian density ${\cal H}_{T}$ 
for the pure tachyon is almost a constant, the
energy density (\ref{eden}) can be compared with that in the materials
\begin{equation}
\frac{T_{3}}{2}\left( {\bf D}\cdot{\bf E}+{\bf H}\cdot{\bf B}\right).
\end{equation}
Then this material may be approximated by a dielectric material at least
formally
\begin{equation}
{\bf D}\approx \frac{\bf E}{V(T)^{2}},\qquad {\bf H}\approx {\bf B},
\end{equation}
where its time-dependent dielectric function diverges as time elapses,
$\varepsilon(t)\approx 1/V(T)^{2}
\stackrel{t\rightarrow \infty}{\longrightarrow}\infty$.

\section{Cosmological Aspects of Rolling Tachyon}
An attractable topic of the rolling tachyon is its application 
to cosmology~\cite{Gib,Ale,KL,tacos,KKK} as indicated from the 
beginning~\cite{Sen}. The simplest set for the cosmology of the 
rolling tachyon is composed of the graviton $g_{\mu\nu}$ and the tachyon $T$.
Therefore, the action is read from Eq.~(\ref{act})
\begin{equation}\label{gtac}
S=\int d^4x\sqrt{-g}\left(\frac{1}{2\kappa^2}R-T_{3}V(T)
\sqrt{1+g^{\mu\nu}\partial_{\mu}T\partial_{\nu}T}\right).
\end{equation}

Without any computation, we read a few intriguing subjects from characters of
the tachyon potential in Eqs.~(\ref{V1})--(\ref{V4}) 
(Refer also to Fig.~\ref{fig1}). At initial stage
of the rolling tachyon, there exists a cosmological constant 
$V(T\approx 0)\approx 1$ so that one expects easily solutions of inflationary 
universe. At late-time of $T\rightarrow \infty$, we have tiny nonvanishing 
but monotonically-decreasing cosmological constant at each instant, which
lets us consider it as a possible source of quintessence. Bad news for
comprising a realistic cosmological model may be absence of reheating 
and difficulty of density perturbation at late-time due to disappearance
of perturbative degrees reflecting nonexistence of stationary vacuum point
in the monotonically-decreasing tachyon potential.

For cosmological solutions, we try a spatially-homogeneous but 
time-dependent solution
\begin{equation}
ds^2=-dt^2+a^2(t)d\Omega^{2}_{k},\quad T=T(t),
\label{cosmos1}
\end{equation}
where $d\Omega^{2}_{k}$ corresponds, at least locally, to the metric of
$S^{3}$, $E^{3}$, or $H^{3}$ according to the value of $k=1,0,-1$, respectively.
Then the Einstein equations are summarized as 
\begin{equation}\label{Eeq1}
\frac{\dot a^2}{a^2}+\frac{k}{a^2} = \frac{\kappa^2}{3} 
\frac{T_{3}V(T)}{\sqrt{1-\dot{T}^{2}}},
\end{equation}
\begin{equation}\label{Eeq2}
\frac{\ddot a}{a}=\frac{\kappa^2}{3}\frac{T_{3}V(T)}{\sqrt{1-\dot{T}^{2}}}
\left(1-\frac{3}{2}\dot{T}^{2}\right),
\end{equation}
and the tachyon equation becomes
\begin{equation}\label{Teq}
\frac{\ddot T}{1-\dot T^2}+3\frac{\dot a}{a}\dot T
+\frac{1}{V}\frac{dV}{dT}=0,
\end{equation}
where is equivalent to conservation equation of the energy-momentum tensor.

For the flat universe of $k=0$, cosmological evolution is rather simple
since both sides of Eq.~(\ref{Eeq1}) are positive semi-definite.
The second equation (\ref{Eeq2}) tells us that the expanding universe 
is accelerating at onset of $T(t)\approx 0$ since the right-hand side of
Eq.~(\ref{Eeq2}) is also positive definite. As time $t$ goes, $\dot{T}$ 
exceeds $\sqrt{2/3}$ so that expansion of the universe slows down and then
finally the scale factor $a(t)$ will halt as 
$a(t)\stackrel{t\rightarrow\infty}{\longrightarrow}{\rm constant}$.
This makes the second term of the tachyon equation (\ref{Teq}) vanish
at infinite time so that we confirm $\ddot{T}\rightarrow 0$ and
$\dot{T}\rightarrow 1$ as $T\rightarrow \infty$.

\section{String Cosmology with Rolling Tachyon}
In this section, we study the role of rolling tachyons in the cosmological 
model with dilatonic gravity. In the string frame, flat space solutions of 
both initial-stage and late-time will be obtained in closed form.
In the Einstein frame, we will show that every expanding solution in flat space
is decelerating.

\subsection{String frame}\label{subsec:sfr}
We begin with a cosmological model induced from string theory, which
is confined on a D3-brane and
includes graviton $g_{\mu\nu}$, dilaton $\Phi$, and tachyon $T$.
In the string frame, the effective action of the bosonic sector of the D3-brane
system is given by 
\begin{eqnarray}
S&=&\frac{1}{2\kappa^2}\int d^4x\sqrt{-g}\;e^{-2\Phi}\left(R
+4\nabla_\mu\Phi\nabla^\mu\Phi\right)
\nonumber\\
&&-T_{3}\int d^4x\;e^{-\Phi}\;V(T)
\sqrt{-\det (g_{\mu\nu}+\partial_{\mu}T\partial_{\nu}T)}\; ,
\label{act2}
\end{eqnarray}
where we turned off an Abelian gauge field $A_{\mu}$ on the D3-brane and
antisymmetric tensor fields of second rank $B_{\mu\nu}$ both on the brane and
in the bulk.
Equations of motion for the metric, the dilaton, and the tachyon
derived from the action (\ref{act2}) are
\begin{eqnarray}\label{metric1}
R_{\mu\nu}-\frac12g_{\mu\nu}R =
\kappa^2(T^{\Phi}_{\mu\nu}+e^{\Phi}T^{T}_{\mu\nu}),
\end{eqnarray}
\begin{equation}
4\left(\nabla_\mu\nabla^\mu\Phi-\nabla_\mu\Phi \nabla^\mu\Phi\right)+R
= \kappa^2T_3\;e^\Phi\;V(T)\sqrt{1+\nabla_\alpha T\nabla^\alpha T},
\label{dilaton1}
\end{equation}
\begin{equation}
\nabla_\mu\nabla^\mu T
-\frac{\nabla_\mu \nabla_\nu T\nabla^\mu T\nabla^\nu T}
 {1+\nabla_\alpha T\nabla^\alpha T}
-\nabla_\mu T\nabla^\mu\Phi
-\frac{1}{V}\frac{dV}{dT}=0,
\label{tachyon1}
\end{equation}
where energy-momentum tensor of the dilaton is
\begin{equation}\label{tmnd}
T^{\Phi}_{\mu\nu}=-\frac{2}{\kappa^{2}}
\left[\nabla_\mu\nabla_\nu\Phi
+g_{\mu\nu}\left(\nabla_\alpha\Phi\nabla^\alpha\Phi
-\nabla_\alpha\nabla^\alpha\Phi\right)\right],
\end{equation}
and that of the tachyon
\begin{equation}\label{tmnt}
T^{T}_{\mu\nu}=
T_3 \frac{V(T)}{\sqrt{1+\nabla_\alpha T\nabla^\alpha T}}\left[
\nabla_\mu T\nabla_\nu T
-g_{\mu\nu}(1+\nabla_\alpha T\nabla^\alpha T) \right].
\end{equation}

For cosmological solutions in the string frame, we try in addition to 
the metric ansatz (\ref{cosmos1})
\begin{equation}\label{Pt}
\Phi=\Phi(t).
\end{equation}
{}From the action (\ref{act2}),
we obtain the following equations
\begin{equation}
3\left(\frac{\dot a^2}{a^2}+\frac{k}{a^2}\right)
-2\left(3\frac{\dot a}{a}\dot\Phi-\dot\Phi^2\right) = 
\kappa^2 e^\Phi \rho_{T},
\label{metric3-1}
\end{equation}
\begin{equation}
2\frac{\ddot a}{a}+\frac{\dot a^2}{a^2}+\frac{k}{a^2}
-2\left(\ddot\Phi+2\frac{\dot a}{a}\dot\Phi-\dot\Phi^2\right) = 
-\kappa^2 e^\Phi p_{T},
\label{metric3-2}
\end{equation}
\begin{equation}
4\left(\ddot\Phi+3\frac{\dot a}{a}\dot\Phi-\dot\Phi^2\right)
-6\left(\frac{\ddot a}{a}+\frac{\dot a^2}{a^2}+\frac{k}{a^2}\right) = 
\kappa^2 e^\Phi p_{T},
\label{dilaton3}
\end{equation}
\begin{equation}
\frac{\ddot T}{1-\dot T^2}+\left(3\frac{\dot a}{a}-\dot\Phi\right)\dot T
+\frac{1}{V}\frac{dV}{dT}=0,
\label{tachyon3}
\end{equation}
where tachyon energy density $\rho_{T}$ and pressure $p_{T}$
defined by $T_{\;\;\;\;\nu}^{T\mu}\equiv
{\rm diag}(-\rho_{T},p_{T},p_{T},p_{T})$ are
\begin{equation}\label{pre}
\rho_T = T_{3}\frac{V(T)}{\sqrt{1-\dot T^2}}~~~{\rm and}
~~~p_T = -T_{3}V(T)\sqrt{1-\dot T^2},
\end{equation}
which formally coincide with Eq.~(\ref{taeq}) and Eq.~(\ref{press}).
The tachyon equation (\ref{tachyon3}) is equivalent to the following
conservation equation
\begin{equation}
\dot\rho_T+(3H-\dot\Phi)\dot T^2\rho_T=0,
\label{conserv-eq}
\end{equation}
where $H=\dot a/a$ is the Hubble parameter.
In the absence of detailed knowledge of $V(T)$,
we will examine characters of the solutions
based on the simplicity of tachyon equation of state 
\begin{equation}
p_T=w_T\rho_T,\quad w_T=\dot T^2-1,
\label{T-eq-of-state}
\end{equation}
which is exactly the same as Eq.~(\ref{rw1}). 

By defining the shifted dilaton $\phi=2\Phi-3\ln a$,
we rewrite the equations (\ref{metric3-1})--(\ref{tachyon3}) as
\begin{eqnarray}
\dot\phi^2-3H^2 +6\frac{k}{a^{2}}
&=& 2\kappa^{2}e^{\frac{\phi}{2}}a^{\frac{3}{2}}\rho_T, \label{metric4-1} \\
2(\dot{H}-H\dot\phi) +4\frac{k}{a^{2}}
&=& \kappa^{2} e^{\frac{\phi}{2}}a^{\frac{3}{2}} p_T, \label{metric4-2} \\
\dot\phi^2-2\ddot\phi +3H^2 +6\frac{k}{a^{2}}
&=& -\kappa^{2} e^{\frac{\phi}{2}}a^{\frac{3}{2}} p_T, \label{dilaton4} \\
\frac{\ddot T}{1-\dot T^2}+\frac12 (3H-\dot\phi)\dot T
&=&-\frac{1}{V}\frac{dV}{dT} .
\label{tachyon4}
\end{eqnarray}
Note that $\sqrt{-g}$, or $a^3$ is not a scalar quantity
even in flat spatial geometry,
the shifted dilaton $\phi$ is not a scalar field in 3+1 dimensions.
The conservation equation (\ref{conserv-eq}) becomes
\begin{equation} \label{conserv-eq2}
\dot\rho_T + \frac12 (3H-\dot\phi)\dot T^2 \rho_T=0.
\end{equation}
Now Eqs.~(\ref{metric4-1})--(\ref{dilaton4}) and Eq.~(\ref{T-eq-of-state})
are summarized by the following two equations
\begin{eqnarray}
2\ddot\phi-\dot\phi^2+2H\dot\phi-3H^2-2\dot H &=& -10\frac{k}{a^{2}},
\label{xx1} \\
w_T\dot\phi^2+4H\dot\phi-3w_TH^2-4\dot H &=& -(8+6w_{T})\frac{k}{a^{2}}.
\label{xx2}
\end{eqnarray}
Let us consider only the flat metric $(k=0)$ in the rest part of the paper.
If we express the dilaton $\phi$ as a function of the scale factor $a(t)$,
$\phi(t)=\phi(a(t))$, we can introduce a new variable $\psi$ such as
\begin{equation}
\psi\equiv a\phi' =\frac{\dot{\phi}}{H},
\end{equation}
where the prime denotes the differentiation with respect to $a$, and the
second equality shows that $\psi$ is the ratio between $\dot{\phi}$ and $H$.
Then Eqs.~(\ref{xx1}) and (\ref{xx2}) are combined into a single first-order
differential equation for $\psi$ :
\begin{equation}\label{psi-eq}
4a\psi'+(\psi^2-3)(w_T\psi+2-w_T) = 0. 
\end{equation}

{}From now on we look for the solutions of Eq.~(\ref{psi-eq}).
Above all one may easily find a constant solution
$\psi=\mp\sqrt{3}$ which is consistent with
Eqs.~(\ref{metric4-1})--(\ref{dilaton4}) only when $\rho_T=0$~:
\begin{eqnarray}
a(t)&=&a_{0}(1\mp\sqrt{3}H_{0}t)^{\mp 1/\sqrt{3}},
~~~H(t)=\frac{H_{0}}{1\mp\sqrt{3}H_{0}t},\label{ata}\\
\Phi(t)&=&\Phi_{0}-\frac{1\pm\sqrt{3}}{2}\ln(1\mp\sqrt{3}H_{0}t),
\label{pta}
\end{eqnarray}
where $H_{0}=H(t=0)$, $a_{0}=a(t=0)$, and $\Phi_{0}=\Phi(t=0)$
throughout this section. However, exactly-vanishing tachyon density
$\rho_{T}=0$ from Eq.~(\ref{metric4-1}) restricts
strictly the validity range of this
particular solution to that of vanishing tachyon potential, $V(T)=0$,
which leads to $T=\infty$.
The tachyon equation (\ref{tachyon4}) forces $\ddot{T}=0$ and $\dot{T}=1$ so
that the tachyon decouples $(w_{T}=p_{T}=0)$.
Therefore, the obtained solution (\ref{ata})--(\ref{pta}) corresponds to
that of string cosmology of the graviton and the dilaton before stabilization
but without the tachyon.

Since it is difficult to solve Eq.~(\ref{psi-eq}) with dynamical $w_{T}$,
let us assume that $w_T$ is time-independent
(or equivalently $a(t)$-independent).
We can think of the cases where
the constant $w_T$ can be a good approximation.
{}From the tachyon potential (\ref{V3}),
the first case is onset of tachyon rolling around the maximum point
and the second case is late-time rolling at large $T$ region.
In fact we can demonstrate that those two cases are the only possibility
as far as no singularity evolves.

When $w_T$ is a nonzero constant, Eq.~(\ref{psi-eq}) allows
a particular solution
\begin{equation}
\psi=\frac{w_T-2}{w_T}\equiv\beta.
\end{equation}
This provides a consistent solution of Eqs.~(\ref{metric4-1})--(\ref{dilaton4})
\begin{eqnarray}
a(t) &=& a_0\left(1+\frac{w_T^2+2}{2w_T}H_0t\right)^{\frac{2w_T}{w_T^2+2}},
\qquad
H(t) =H_0\left(1+\frac{w_T^2+2}{2w_T}H_0t\right)^{-1}, 
\\
\Phi(t)&=&\Phi_{0}+\frac{2(2w_{T}-1)}{w^{2}_{T}+2}\ln
\left(1+\frac{w^{2}_{T}+2}{2w_{T}}H_{0}t\right). 
\label{ptw}
\end{eqnarray}
{}From Eq.~(\ref{metric4-1}) and Eq.~(\ref{conserv-eq2}),
the tachyon energy density $\rho_{T}$ is given by
\begin{eqnarray}\label{rhos}
\rho_T(t) = \frac{2-2w_T-w_T^2}{w_T^2\kappa^2e^{\Phi_0}}H_{0}^{2}
\left(1+\frac{w_T^2+2}{2w_T}H_0 t\right)^{-\frac{2(1+w_T)^2}{w_T^2+2}} .
\end{eqnarray}
Since the obtained solution is a constant solution of $\psi$,
it has only three initially-undetermined constants.
Specifically, the solution should satisfy $\dot\Phi=[(2w_T-1)/w_T]H$
so that the initial conditions also satisfy a relation
$\dot\Phi_0=[(2w_T-1)/w_T]H_0$.
Once we assume general solutions of $a(t)$-dependent $\psi$
with keeping constant nonzero $w_{T}$, they should be classified 
by four independent parameters $(a_{0},H_{0},\Phi_{0},\dot{\Phi}_{0})$
instead of three in Eq.~(\ref{ptw}).

According to the aforementioned condition for valid $w_{T}$ values,
the obtained solution in Eq.~(\ref{ptw}) may be physically
relevant as the onset solution of $w_{T}=-1$ $(\psi=3)$.
In this case, $\rho_T(t)$ is reduced to a constant
$ \rho_T(t) = 3 e^{-\Phi_0} H_0^2 /\kappa^2 $.
Comparing this with the definition of $\rho_T$ in Eq.~(\ref{pre}),
the initial Hubble parameter $H_0$ is related to the dilaton as
$ H_0 = \pm \kappa e^{\Phi_0/2} \sqrt{T_3/3} $. Then, with $T(t) = 0$,
the tachyon equation of motion is automatically satisfied and hence
Eq.~(\ref{ptw}) becomes an exact solution of the
whole set of equations of motion (\ref{metric3-1})--(\ref{tachyon3}).
Since the tachyon field remains as constant at the maximum of the potential,
this solution describes the expanding or shrinking solution depending
on the initial Hubble parameter, with a constant vacuum energy corresponding
to brane tension due to tachyon sitting at the unstable
equilibrium point.\footnote{The interpretation as expanding or
shrinking solution needs to be more careful,
since we are working in the string frame.
Actually the behaviors are reversed in the Einstein frame as we will
see in the next subsection.}

In order to study the behavior of the tachyon rolling down from the top of
the potential, now we slightly perturb this solution, i.e., look for a
solution with nonzero but small $T$ dependence.
So we treat $T$ as a small expansion parameter
and work up to the first-order in $T$.
Since the unperturbed solution satisfies $ 3H = \dot\Phi $,
the tachyon equation of motion (\ref{tachyon3}) becomes, to the
first-order in $T$,
\begin{equation}
\ddot{T} = -\frac{1}{V} \frac{dV}{dT} .
\end{equation}
This can easily be integrated to
\begin{equation}\label{teqn}
\frac12\dot{T}^2 = -\ln V + \mbox{const} = \frac{T^2}{8\ln2} + \mbox{const},
\end{equation}
where we used the form of the potential near the origin (\ref{V3}).
Given the initial condition that $T=T_0$ and $\dot T = 0$ at $t=0$,
we can solve this equation (\ref{teqn}) and obtain
\begin{equation} \label{perturbed}
T(t) = T_0 \cosh{\alpha t},
\end{equation}
where $\alpha = 1/2\sqrt{\ln2}$.
Therefore tachyon starts to roll down the potential as a hyperbolic
cosine function.
Taking derivative, we find $\dot T = \alpha T_0 \sinh{\alpha t}$.
The range for which $\dot T$ remains small is then
$t\lesssim t_r\equiv 2\sqrt{\ln2} \sinh^{-1}(2\sqrt{\ln2}/T_0)$,
during which the approximation $w_T \simeq -1$ is good.
Unless the initial value $T_0$ is fine-tuned,
the tachyon follows the onset solution (\ref{ptw})--(\ref{rhos})
for $t\lesssim t_r$ and enters into rolling mode.

For more general solutions, the first-order differential
equation (\ref{psi-eq}) can be integrated to
\begin{equation}
a = C\left[\frac{\psi^2-3}{(\psi-\beta)^2}
\left(\frac{\psi-\sqrt3}{\psi+\sqrt3}\right)^{\frac{\beta}{\sqrt3}}
\right]^{\frac{\beta-1}{\beta^2-3}},
\end{equation}
where $C$ is an integration constant.
Note that this algebraic equation does not provide a closed form of $\psi$
in terms of the scale factor $a(t)$ except for a few cases, e.g.,
$w_T=0,-1/(\sqrt{3}-1/2),-2/(3\sqrt{3}-1)$.

Fortunately, for the late-time case of vanishing $w_T$,
we can obtain the solution in closed form
\begin{eqnarray}\label{cpm}
\psi = \sqrt3\ \frac
{C_+\left(\frac{a}{a_0}\right)^{\sqrt3/2}
+C_-\left(\frac{a}{a_0}\right)^{-\sqrt3/2}}
{C_+\left(\frac{a}{a_0}\right)^{\sqrt3/2}
-C_-\left(\frac{a}{a_0}\right)^{-\sqrt3/2}}.
\end{eqnarray}
Then the scale factor $a$ and the dilaton $\Phi$ are explicitly
expressed as functions of time $t$ by solving the equations
(\ref{metric4-2})--(\ref{dilaton4}):
\begin{eqnarray}
a(t) &=& a_0\left(\frac{C_-t+2}{C_+t+2}\right)^{1/\sqrt3},
\qquad H(t)=\frac{4H_0}{(C_-t+2)(C_+t+2)},
\label{att}\\
\Phi(t)&=&\Phi_0+\ln\left[
2\frac{(C_-t+2)^{(\sqrt3-1)/2}}{(C_+t+2)^{(\sqrt3+1)/2}}
\right],
\label{a-w0}
\end{eqnarray}
where $C_\pm=(3\mp\sqrt3)H_0-2\dot\Phi_0$.
We also read the tachyon density $\rho_T$ from Eq.~(\ref{metric4-1})
\begin{equation}\label{rhot}
\rho_T = C_+C_-e^{-\Phi_0}\ 
   \frac{(C_+t+2)^{(\sqrt3-1)/2}}{(C_-t+2)^{(\sqrt3+1)/2}}.
\end{equation}

Note that $C_{\pm}$ should have the same sign
from the positivity of the tachyon density (\ref{rhot}).
Let us first consider that both $C_{+}$ and $C_{-}$ are positive.
When $C_->C_+$ or equivalently $H_0>0$, the scale factor $a$ is growing
but saturates to a finite value
such as $a(\infty)=a_0 (C_-/C_+)^{1/\sqrt{3}}$ in the string frame.
When $C_-<C_+$, it decreases.
When $C_-=C_+$, $H_0=0$ so that the scale factor is a constant, $a(t)=a_0$.
For all of the cases, the dilaton $\Phi$ approaches negative infinity.
Note that $w_{T}=0$ means late-time, the tachyon density decreases like
$\rho_T\sim1/t$ as $t\rightarrow\infty$. Consistency check by using
Eq.~(\ref{conserv-eq2}) or equivalently by Eq.~(\ref{tachyon4}) provides us
the expected result, $\dot{T}\rightarrow1$.
If both $C_{+}$ and $C_{-}$ are negative, there appears a singularity at
finite time irrespective of relative magnitude of $C_{+}$ and $C_{-}$.

\subsection{Einstein frame}\label{subsec:Efr}
In the previous subsection, it was possible to obtain the cosmological solutions
analytically in a few simple but physically meaningful limiting cases.
To study the physical implications of what we found,
however, we need to work in the Einstein frame.
In this subsection, we will convert the cosmological solutions obtained
in the string frame to those in the Einstein frame
and discuss the physical behaviors.

First of all, let us rewrite the action (\ref{act2}) in the Einstein frame 
of which metric
$g^E_{\mu\nu}$ is related to the string metric by a conformal transformation
\begin{equation}
g_{\mu\nu}=e^{2\Phi}g^E_{\mu\nu}.
\end{equation}
Note that we abbreviate the superscript $E$ for convenience in what follows.
Then the action (\ref{act2}) is changed to
\begin{eqnarray}
S&=&\frac{1}{2\kappa^2}\int d^4x\sqrt{-g}\left(R
-2\nabla_\mu\Phi\nabla^\mu\Phi\right)
\nonumber\\
&&-T_{3}\int d^4x\sqrt{-g}\;e^{3\Phi}\;V(T)
\sqrt{1+e^{-2\Phi}\nabla_\mu T\nabla^\mu T}.
\label{act4}
\end{eqnarray}
Equations of motion are
\begin{eqnarray}\label{metric9}
R_{\mu\nu}-\frac12g_{\mu\nu}R =\kappa^{2}(T^{\Phi}_{\mu\nu}+
e^{3\Phi}T^{T}_{\mu\nu}),
\end{eqnarray}
\begin{equation}
\nabla_\mu\nabla^\mu\Phi =
\frac{\kappa^2}{2}T_3\;e^{3\Phi}V(T)
\frac{3+2e^{-2\Phi}\nabla_\alpha T\nabla^\alpha T}{
\sqrt{1+e^{-2\Phi}\nabla_\alpha T\nabla^\alpha T}},
\label{dilaton9}
\end{equation}
\begin{equation}
\nabla_\mu\nabla^\mu T
+\nabla_\mu T\nabla^\mu\Phi
-\frac{e^{-2\Phi}\left(\nabla_\mu\nabla_\nu T-\nabla_\mu\Phi\nabla_\nu T\right)
\nabla^\mu T\nabla^\nu T}
{1+e^{-2\Phi}\nabla_\alpha T\nabla^\alpha T}
-\frac{e^{2\Phi}}{V}\frac{dV}{dT}=0,
\label{tachyon9}
\end{equation}
where the corresponding energy-momentum tensors are
\begin{equation}\label{etmnp}
T^{\Phi}_{\mu\nu}=\frac{1}{\kappa^{2}}\left(2\nabla_\mu\Phi\nabla_\nu\Phi
-g_{\mu\nu}\nabla_\alpha\Phi\nabla^\alpha\Phi\right),
\end{equation}
\begin{equation}\label{etmnt}
T^{T}_{\mu\nu}=T_{3}
\frac{V(T)}{\sqrt{1+e^{-2\Phi}\nabla_\alpha T\nabla^\alpha T}}
\left[ e^{-2\Phi}\nabla_\mu T\nabla_\nu T
-g_{\mu\nu}(1+e^{-2\Phi}\nabla_\alpha T\nabla^\alpha T)
\right].
\end{equation}

For cosmological solutions, we use again Eq.~(\ref{cosmos1}) and 
Eq.~(\ref{Pt})
and then the equations of motion (\ref{dilaton9})--(\ref{tachyon9}) become
\begin{equation} \label{einstein1}
\frac{\dot a^2}{a^2}+\frac{k}{a^2}
=\frac{\kappa^{2}}{3}(\rho_{\Phi}+e^{3\Phi}\rho_{T})
= \frac{1}{3}\dot\Phi^2
+\frac{\kappa^2}{3}T_3\;e^{3\Phi}V(T)\frac{1}{\sqrt{1-e^{-2\Phi}\dot T^2}},
\end{equation}
\begin{eqnarray} 
\frac{\ddot a}{a}
&=&-\frac{\kappa^{2}}{6}\left[(\rho_{\Phi}+3p_{\Phi})
+e^{3\Phi}(\rho_{T}+3p_{T}) \right] \nonumber\\
&=&-\frac{2}{3}\dot\Phi^2
+\frac{\kappa^{2}}{6}T_{3}\;e^{3\Phi}V(T)
\frac{2-3e^{-2\Phi}\dot T^2}{\sqrt{1-e^{-2\Phi}\dot T^2}},
\label{einstein2}
\end{eqnarray}
\begin{eqnarray} \label{einstein3}
\ddot\Phi+3\frac{\dot a}{a}\dot\Phi 
= -\frac{\kappa^2}{2}e^{3\Phi}(\rho_{T}-2p_{T}) 
= -\frac{\kappa^2}{2}T_3\;e^{3\Phi}V(T)
\frac{3-2e^{-2\Phi}\dot T^2}{\sqrt{1-e^{-2\Phi}\dot T^2}},
\end{eqnarray}
\begin{equation} \label{einstein4}
\frac{\ddot T}{1-e^{-2\Phi}\dot T^2}
+3H\dot T 
+\dot\Phi\dot T \frac{1-2e^{-2\Phi}\dot T^2}{1-e^{-2\Phi}\dot T^2}
+e^{2\Phi}\frac{1}{V}\frac{dV}{dT}=0.
\end{equation}
Tachyon energy density $\rho_T$ and pressure $p_T$ in the Einstein frame
are obtained by the replacement $\dot{T}_{s}=e^{-\Phi}\dot{T}$
in Eq.~(\ref{pre}),
\begin{equation}\label{rpt}
\rho_T = T_{3}\frac{V(T)}{\sqrt{1-e^{-2\Phi}\dot T^2}}~~~{\rm and}
~~~p_T = -T_{3}V(T)\sqrt{1-e^{-2\Phi}\dot T^2} ,
\end{equation}
and thereby $w_{T}$ is given as
\begin{equation}
w_{T}\equiv p_T /\rho_T=e^{-2\Phi}\dot T^2 -1.
\end{equation}

Alternatively, we can obtain the same equations
from Eqs.~(\ref{metric3-1})--(\ref{tachyon3})
by using the metric of the form~\footnote{In this subsection 
all the quantities are in the Einstein frame
except the variables with subscript $s$ which denote the quantities
in the string frame.}
\begin{equation} \label{einsteinmetric}
ds^2=e^{2\Phi}(-dt^2+a^2(t)d\Omega^{2}_{k}),
\end{equation}
and hence the time $t$ and the scale factor $a$ are related to
those in the string frame as
\begin{equation}\label{rel}
a_s=ae^{\Phi},\qquad dt_s=e^{\Phi}dt.
\end{equation}
Then the Einstein equations for the flat
case $(k=0)$ in the string frame (\ref{metric3-1})--(\ref{tachyon3})
are converted to
\begin{eqnarray}
H^2 &=& \frac13 \dot\Phi^2 + \frac13 \kappa^2 e^{3\Phi}\rho_T\,, 
           \label{einstein5}\\
\dot H &=& - \dot\Phi^2 - \frac12 \kappa^2 e^{\Phi}\rho_T \dot T^2\,. 
           \label{einstein6}
\end{eqnarray}
Demanding constant $w_{T}$ is nothing but asking a strong
proportionality condition between the dilaton and tachyon,
$\dot{T}\propto e^{\Phi}$.
Note that the pressure $p_{T}$ as shown in Eq.~(\ref{pre}) is always
negative irrespective
of both specific form of the tachyon potential $(V(T)\ge 0)$ and the value of
the kinetic term $(e^{-2\Phi}\dot{T}^{2}\le 1)$, and the value of $w_T$
interpolates smoothly between $-1$ and 0.

First we observe that the right-hand side of Eq.~(\ref{einstein5}) is always
positive, which means that the Hubble parameter $H(t)$ is either
positive definite or negative definite for all $t$ and it cannot change
the sign in the Einstein frame.
Obviously it is a natural consequence of the weak energy condition.
Let us first consider the case of positive Hubble parameter, $H(t)>0$.
Eq.~(\ref{einstein6}) shows $\dot H$ consists of two
terms both of which are negative definite for all $t$.  Since $H>0$ by
assumption, the only consistent behavior of $H$ in this case is that
$\dot H$ vanishes as $t\rightarrow \infty$, which, in turn implies that
$\dot\Phi$ and $e^{\Phi}\rho_T \dot T^2$ go to zero, separately.
It also means that $H$ should be a regular function for all $t$. In order to
find the large $t$ behavior of $H(t)$, one has to study $e^{-\Phi} \dot T$
in large $t$ limit which appears in the definition of $w_T$ in the
Einstein frame. Knowing that the functions are regular, it is not difficult
to show that the only possible behavior is $e^{-\Phi} \dot T \rightarrow 1$
as $t \rightarrow \infty$ after some straightforward analysis of
Eqs.~(\ref{einstein5})--(\ref{einstein6}). Combining it with the fact that
$\dot\Phi$ and $e^{\Phi}\rho_T \dot T^2$ vanish, we can immediately
conclude from Eq.~(\ref{einstein5}) that $H(t)$ should go to zero in large
$t$ limit.

The asymptotic behavior of fields in case of the positive Hubble parameter
can be found from the solution (\ref{a-w0}) since $w_T$ is essentially
zero for large $t$ as we just have seen above. The only thing to do is
to transform the expressions in the string frame to those in the
Einstein frame, using the relation (\ref{rel}).
Therefore, for large $t$, we find
\begin{eqnarray} \label{einstein_a}
a(t_s) &=& a_s(t_s)e^{-\Phi(t_s)}
\simeq\frac{1}{2}a_{s0}e^{-\Phi_{0}}
(C_{+}t_s+2)^{\frac{\sqrt3+1}{2\sqrt3}}
(C_{-}t_s+2)^{\frac{\sqrt3-1}{2\sqrt3}}, \nonumber \\
t &=& \int dt_s e^{-\Phi} 
    \simeq 2 e^{-\Phi_0} \int dt_s 
       \frac{(C_+ t_s+2)^{(\sqrt3+1)/2}}{(C_- t_s+2)^{(\sqrt3-1)/2}}.
\end{eqnarray}
One can also identify the initial Hubble parameter $H_0$ in terms of $C_\pm$ as
\begin{equation}
H_0 = \frac14 e^{\Phi_0}\left[
               \left( 1 - \frac1{\sqrt3} \right) C_{-} 
             + \left( 1 + \frac1{\sqrt3} \right) C_{+} \right].
\end{equation}
Note that $C_\pm$ have the same sign as the Hubble parameter $H$.
Now, with $C_\pm>0$, one can easily confirm from the above equation
(\ref{einstein_a}) that all the functions indeed behave regularly.
In $t_s\rightarrow\infty$ limit, $a\sim t_s$ and $t \sim t_s^{2}$
so that the asymptotic behavior of the scale factor becomes $a\sim t^{1/2}$.
This power law expansion in flat space is contrasted with the result of
Einstein gravity without the dilaton $\Phi$, where ultimately the scale
factor ceases to increase, $\lim_{t\rightarrow\infty}a(t)\rightarrow$ constant.
The behavior of tachyon density $\rho_T$ can be read from
Eq.~(\ref{rhot}) with $t$ replaced by $t_s$, which shows that
$\rho_T \sim t^{-1/2}$.
Since $w_T$ also goes to zero, the fluid of condensed tachyon becomes
pressureless.
Differently from ordinary scalar matter where matter domination
of pressureless gas is achieved
for the minimum kinetic energy $(\dot{T}\rightarrow 0)$, it is
attained for the maximum value of time dependence $(e^{-\Phi}\dot{T}
\rightarrow 1$ as $T\rightarrow \infty)$ for the tachyon potential
given in Eq.~(\ref{V3}).

When the Hubble parameter $H$ is negative, the situation is a bit more
complicated. Since $\dot H<0$ always,
$H$ becomes more and more negative and there is a possibility that
eventually $H$ diverges to negative infinity at some finite time.
Indeed, it turns out that all solutions in this case develop a singularity
at some finite time at which $H \rightarrow -\infty$ and $a \rightarrow 0$.
These big crunch solutions may not describe viable universes
in the sense of observed cosmological data.
Depending on initial conditions, the dilaton $\Phi$ diverges to either $\infty$ or
$-\infty$ and $\dot T$ goes to either $\infty$ or zero with the factor
$e^{-\Phi} \dot T$ remaining finite. It is rather tedious and not much
illuminating to show this explicitly, so here we will just content
ourselves to present a simple argument to understand the behavior.
Since the tachyon field $T$ rolls down from the maximum of the potential
to the minimum at infinite $T$, it is physically clear
that $ \dot T_s = e^{-\Phi} \dot T$ would eventually go to one unless
there is a singularity at some finite time. Suppose that there appeared
no singularity until some long time had passed so that $e^{-\Phi} \dot T$
approached to one sufficiently closely. Then Eq.~(\ref{einstein_a}) should be
a good approximate solution in this case. However, we know that both
$C_\pm$ are negative when $H<0$ and Eq.~(\ref{einstein_a}) is clearly
singular in this case. We have also verified the singular behavior for
various initial conditions using numerical analysis.

As mentioned in the previous section, the tachyon $T$ is decoupled when
$e^{-\Phi}\dot{T}=1$ and $T=\infty$. In this decoupling limit, characters of
the Einstein equations (\ref{einstein5})--(\ref{einstein6}) that $H^{2}>0$
and $\dot{H}<0$ do not change so that all the previous arguments can be applied.
Well-known cosmological solution of the dilaton gravity before stabilization
of the dilaton is
\begin{eqnarray}
a(t)&=&a_0 (1+3H_{0}t)^{1/3},~~~H(t)=\frac{H_{0}}{1+3H_{0}t},
\label{adg}\\
\Phi(t)&=&\Phi_{0}\pm\frac{1}{\sqrt{3}}\ln(1+3H_{0}t),
\label{pdg}
\end{eqnarray}
where the $(\pm)$ sign in Eq.~(\ref{pdg}) is due to the reflection symmetry
$(\Phi\leftrightarrow -\Phi)$ in the equations
(\ref{einstein5})--(\ref{einstein6}).
This solution can also be obtained throughout a transformation (\ref{rel})
from Eqs.~(\ref{ata})--(\ref{pta}).
For $H_{0}<0$, it is a big crunch solution $(a\rightarrow 0)$ which encounters
singularity $(H\rightarrow\infty,~\Phi\rightarrow \mp\infty)$ as $t\rightarrow
1/3|H_0|$. For $H_{0}>0$, it is an expanding but decelerating solution.
Since $a\sim t^{1/3}$, the power of expansion rate is increased from $1/3$ to
$1/2$ by the tachyonic effect as expected.

So far we discussed generic properties and asymptotic behaviors of solutions
in the Einstein frame. Now we consider the behavior at the onset.
The solution (\ref{ptw}) obtained by assuming constant $w_T$ is
transformed to the Einstein frame as
\begin{eqnarray}
a(t) &=& a_0\left[1+\frac{(w_T-2)^2}{2(1-w_T)}H_0t
        \right]^{\frac{2(1-w_T)}{(w_T-2)^2}}, \nonumber \\
e^{\Phi(t)} &=& e^{\Phi_0}\left[1+\frac{(w_T-2)^2}{2(1-w_T)}H_0t
        \right]^{\frac{2(2w_T-1)}{(w_T-2)^2}},
\end{eqnarray}
where the initial Hubble parameter $H_0$ is related to that in the string
frame by $H_0=e^{\Phi_0}H_{0s}(1-w_T)/w_T$.
Note that $H_0$ and $H_{0s}$ have opposite signs since $w_T<0$. Therefore
the expanding (shrinking) solution in the string frame corresponds to the
shrinking (expanding) solution in the Einstein frame.
For the onset solution with $w_T=-1$, the tachyon energy density $\rho_T$
is a constant as before, $\rho_T(t) = 3e^{-3\Phi_0} H_0^2/4 \kappa^2$.
Then the initial Hubble parameter is given by
$H_0 = \pm 2\kappa e^{3\Phi_0/2} \sqrt{T_3/3}$, which describes the exact
solution that tachyon remains at the origin as explained in section 2.
Under a small perturbation tachyon starts
rolling down according to Eq.~(\ref{perturbed}) with $t$ replaced by
$t_s$. The rest of the discussion on the rolling behavior is the same as
in the string frame and the details will not be repeated here.

In conclusion the cosmological solution can be classified into two categories
depending on the value of the Hubble parameter $H(t)$ in the Einstein frame.
When the initial Hubble parameter $H_0$ is positive, the solution is
regular and the universe is expanding but decelerating as
$a(t) \sim \sqrt{t}$ while $e^{\Phi(t)}$ vanishes. When $H_0$ is negative,
there appears a singularity at some finite time $t$ at which the universe
shrinks to zero.

\section{Summary}

In this paper we have discussed basic field theoretic properties of
rolling tachyon and Abelian gauge field, and then cosmological solutions.
We have shown that both tachyon and Abelian gauge field cannot propagate
in terms of perturbative modes when the homogeneous tachyon rolled
to the infinity which corresponds to the minimum of its potential. 
Throughout the study of homogeneous and isotropic universes in string frame, 
we found initial-stage and late-time solutions in closed form in addition to
the known solution of the gravitons and dilaton in the decoupling
limit of the tachyon. We have also provided a description of 
cosmological solutions in
the Einstein frame~: Once the universe starts expanding, then it continues
expanding eternally but decelerating. 

We conclude with the list of intriguing questions for further study.
The nonlocal effective action we dealt with is not valid for large 
value of the tachyon field, so genuine string theory description 
is needed in such regime, including tower of massive string spectra.
Since both the tachyon and Abelian gauge field cannot support
propagating physical degrees in a perturbative way, an intriguing
question may be existence of extended objects in the background of
the rolling tachyon configuration. In relation with tachyon cosmology,
we may raise a number of questions. Stabilization mechanism of dilaton
should be understood and its positive effect to our unsatisfactory
decelerating universe solutions is carefully investigated.
Various topics for the rolling tachyons asked in the Einstein gravity
should be addressed again in the context of dilaton gravity
such as existence of inflationary era, possibility as a source of
quintessence, reheating without oscillating tachyon modes,
cosmological perturbation and structure formation, and so on.
When the Abelian gauge field in concerned, one can ask whether or not
radiation dominated era is possible.

\section*{Acknowledgments}
Three of the authors(C.Kim, H.B. Kim, and Y. Kim) never forget 
Pong Youl Pac as our teacher and mentor.
Y. Kim would like to acknowledge the hospitality of E-Ken researchers
of Nagoya University where a part of this work has been done.
This work was supported by Korea Research Foundation Grant
KRF-2002-070-C00025(C.K.) and the Swiss Science Foundation,
grant 21-58947.99(H.B.K.), and is
the result of research activities (Astrophysical Research
Center for the Structure and Evolution of the Cosmos (ARCSEC) and
the Basic Research Program, R01-2000-000-00021-0)
supported by Korea Science $\&$ Engineering Foundation(Y.K. and O.K.).

\end{document}